\documentclass[%
reprint,
nobibnotes,
amsmath,amssymb,
prl,
]{revtex4-2}

\usepackage{graphicx}
\usepackage{dcolumn}
\usepackage{bm}
\usepackage{array}
\newcolumntype{M}[1]{>{\centering\arraybackslash}m{#1}}
\usepackage{caption}
\usepackage{subcaption}
\usepackage{float}
\usepackage[dvipsnames]{xcolor}
\definecolor{firetruck}{RGB}{206,32,41}
\definecolor{sapphire}{RGB}{15,82,186}

\begin{document}

\preprint{APS/123-QED}

\title{Strouhal number universality in high-speed cylinder wake flows}

\author{Premika S. Thasu}
\author{Subrahmanyam Duvvuri}%
 \email{Email address: subrahmanyam@iisc.ac.in}
\affiliation{%
 Turbulent Shear Flow Physics and Engineering Laboratory\\
 Department of Aerospace Engineering\\
 Indian Institute of Science\\
 Bengaluru 560012, India
}%




\date{\today}

\begin{abstract}
Flow oscillations in the near-wake region of a 2D circular cylinder are experimentally investigated at Mach 6 over the Reynolds number range $2.3\times10^5$ to $5\times10^5$. The oscillation frequency is obtained by spectral proper orthogonal decomposition of high-speed schlieren data. The Strouhal number based on the length of the near-wake shear layers is found to exhibit universal behavior. This corroborates experimental findings at Mach 4 from recent literature, and further, the universal behavior is also seen to hold with respect to Mach number. Time-resolved pressure measurements at the flow separation points on the cylinder aft surface show that coherent oscillatory activity occurs with a phase difference of $\pi$ radians between the two statistically-symmetric halves of the flow. This aspect of the flow dynamics at high speeds is in common with its low-speed counterpart, \emph{i.e.} the canonical problem of cylinder wake in an incompressible flow.
\end{abstract}

\maketitle

\section{INTRODUCTION \label{sec:intro}}

Flow over a 2D circular cylinder is a classic fluid dynamics problem. In the low-speed flow (incompressible) regime, unsteadiness in the near- and far-wake regions is primarily characterized by shedding of concentrated vorticity from the cylinder surface and vorticity growth driven by convective instability, respectively. An extensive body of literature is available on various aspects pertaining to flow unsteadiness in the low-speed regime \cite[see for instance refs.][and other references cited therein]{roshko1954drag, roshko1961experiments, williamson1996vortex, giannetti2007stability, kumar2012svs}. In comparison, studies of cylinder wakes in the high-speed (supersonic, hypersonic) regime are somewhat limited. Early experimental work detailed the mean (time average) flow structure in the near-wake region of a high-speed cylinder \cite{gowen1953drag, mccarthy1964study, dewey1965near}. More recent studies investigated mean flow features through computations \cite{bashkin2002comparison, park2016base, hinman2017reynolds}.

Unsteady flow features in this canonical wake problem at high speeds have surprisingly received no attention in literature, with the recent work of \citet{schmidt2015oscillations} seemingly being the only exception. Time-resolved shadowgraph visualization of cylinder flow at Mach 4 by \citet{schmidt2015oscillations} revealed coherent oscillations in the near-wake region. Strouhal number behaviour over a decade range in the Reynolds numbers from their experiments suggest the near-wake shear layer length (see figure~\ref{fig:cyl_schem}) to be the governing length scale for the oscillations. Based on the experimental observations a universal value of approximately 0.48 for the Strouhal number was reported. The invariance of Strouhal number with Reynolds number was hypothesized as being indicative of an acoustic resonance mechanism that sustains the oscillations \cite{schmidt2015oscillations}. The current work aims to build on some of these observations toward obtaining a more detailed understanding of unsteady flow features in the near-wake region of a high-speed circular cylinder. Results and analysis from cylinder wake experiments in a Mach 6 flow are presented here.

\section{WIND TUNNEL EXPERIMENTS \label{sec:expts}}

\begin{figure*}[t]
    \centering
    \includegraphics[width=0.75\textwidth]{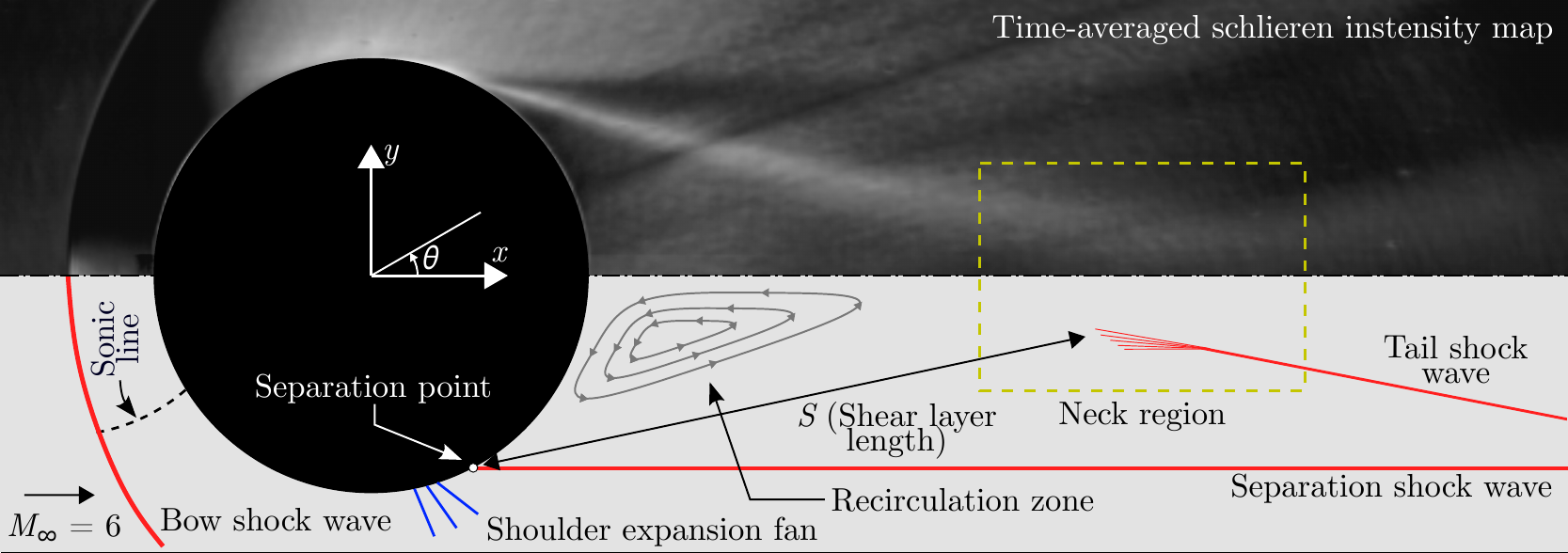}
    \caption{Key features in high-speed flow past a 2D circular cylinder. Top half shows an average schlieren intensity map from run 5 as a representative example, while the bottom half is a qualitative schematic illustration of the flow.}
    \label{fig:cyl_schem}
\end{figure*}

Experiments with 2D circular cylinders at free-stream Mach number $M_{\infty} = 6$ were performed in the Roddam Narasimha Hypersonic Wind Tunnel (RNHWT) at IISc. RNHWT is a pressure-vacuum type 0.5 m diameter enclosed free-jet facility \cite[for more details see ref.][]{sasidharan2021large}. Two cylinder test models of diameters $D$ = 30 and 50 mm and span 120 mm were used. Provisions were made in the 50 mm diameter model to install miniature surface pressure transducers at two locations on the span mid-plane -- $\theta = 62^\circ$ and $\theta = -62^\circ$ (see cylinder-centered coordinate system marked in figure~\ref{fig:cyl_schem}). The choice of $\theta = \pm62^\circ$ is explained later. A similar provision could not be made on the 30 mm diameter cylinder model owing to the size of the pressure transducers. The Reynolds number $Re_D$ with respect to the diameter of the cylinder is defined as
\begin{equation}
    Re_D = \frac{U_{\infty}\,D}{\nu_{\infty}}\, ,
\end{equation}
where $U_{\infty}$ and $\nu_{\infty}$ are free-stream velocity and kinematic viscosity respectively. Experiments were performed at six different Reynolds numbers in the range $2.3\times10^5 \leq Re_D \leq 5\times10^5$; the flow conditions are summarised in table \ref{tab:exp_cond}, where $P_0$ and $T_0$ denoted the total pressure and total temperature, respectively.

\begin{table}[h]
\begin{tabular}{|M{1.1cm}|M{1.1cm}|M{1.1cm}|M{1.1cm}|M{1.1cm}|}
\hline
\hline
Run no.&
$D$ (mm)&
$P_0$ (bar)&
$T_0$ (K)&
$Re_D$ ($\times 10^5$)\\
\hline
1 & 30 & 8.2 & 443 & 2.3 \\
2 & 30 & 9.6 & 432 & 2.8 \\
3 & 30 & 11.1 & 449 & 3.0 \\
4 & 50 & 8 & 430 & 4 \\
5 & 50 & 9.7 & 459 & 4.3 \\
6 & 50 & 10.9 & 445 & 5 \\
\hline
\hline
\end{tabular}
\caption{A summary of flow conditions.}
\label{tab:exp_cond}
\end{table}

In all experiments high-speed schlieren data was recorded at 100,000 frames per second in the near wake flow region; the schlieren set-up is similar to the one used earlier in RNHWT \cite{sasidharan2021large}. The schlieren ``knife-edge'' was set to a horizontal position to capture density ($\rho$) gradients normal to the free-stream flow direction, \emph{i.e.} $\partial \rho\,/\,\partial y$. A large cut-off value for the knife-edge was used to obtain sufficient image sensitivity to density gradients. This introduces an asymmetry between the corresponding bright/dark schlieren bands in the top/bottom halves of the wake, since image light intensity in regions where the light bends towards the knife edge saturates to zero beyond a certain threshold in density gradient \cite{settles2001schlieren}. However, given the physical symmetry, the flow is statistically identical in the top and bottom halves, and no flow information of relevance is lost due to this schlieren artefact. In addition to schlieren, time-resolved pressure measurements were made simultaneously at $\theta = \pm62^\circ$ on the surface of the 50 mm diameter cylinder model (runs 4, 5, 6 in table~\ref{tab:exp_cond}) using flush-mounted piezo-resistive transducers (Kulite XCE-093-5A). These transducers have a sensing diameter of 2.4 mm and resonance frequency of 150 kHz, with the usable bandwidth limited to 30 kHz. The 30 kHz bandwidth, however, easily covers the frequency range of interest in these experiments. Experiments at each of the flow conditions outlined in table~\ref{tab:exp_cond} were performed at least twice to ensure data repeatability.

\section{RESULTS AND DISCUSSION \label{sec:results}}

\begin{figure*}[t]
    \centering
    \includegraphics[width=\textwidth]{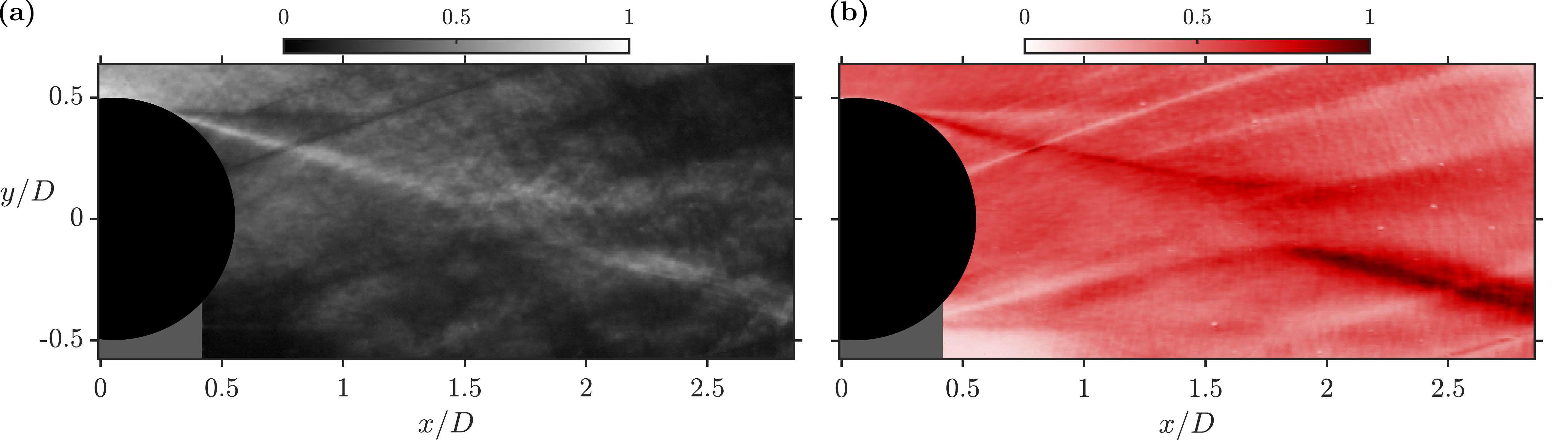}
    \caption{Schlieren data at $Re_D = 4.3\times 10^5$. (a) Instantaneous schlieren snapshot with intensity normalized between 0 and 1. (b) Normalized standard deviation map of temporal fluctuations in schlieren intensity.}
    \label{fig:cyl_inst_std}
\end{figure*}

Flow structure of the high-speed cylinder wake is very different from its low-speed counterpart. Figure~\ref{fig:cyl_schem} illustrates the key features of the mean-flow in the cylinder upstream and near-wake regions. The flow downstream of the leading bow shock wave in the vicinity of the fore stagnation point is subsonic. The fluid from this region accelerates as it moves around the cylinder and goes supersonic beyond the sonic line. A set of expansion fans in the aft region result in a further increase in Mach number as the flow curves around the cylinder. Unlike in an incompressible flow scenario, where an adverse pressure gradient gets established in the aft region, the expanding flow sets up a favourable pressure gradient. While it is natural to expect the cylinder surface boundary layer to not separate under such a condition, separation does occur due to limitations on the turning angle of supersonic flow in the downstream. This leads to the formation of two symmetric supersonic shear layers on either sides of the center line. The shear layers and the cylinder surface enclose two regions of subsonic re-circulating flow with opposing sense of rotation. The region where the two shear layers converge is referred to as the ``neck." Tail shock waves are generated in the neck region which turn the entire flow parallel to the free-stream. In precursor experiments flow separation was found to occur approximately in the range $\theta = \pm[61^\circ, 63^\circ]$ for the $Re_D$ range $4\times10^5 \leq Re_D \leq 5\times10^5$. Hence the two pressure sensors were placed at $\theta = \pm62^\circ$ with the aim of understanding the surface pressure behavior very close to the two separation locations. The key flow features outlined above -- separation shock wave, free shear layers, tail shock wave -- can be identified in the representative schlieren images from run 5 shown in figures~\ref{fig:cyl_schem} and \ref{fig:cyl_inst_std}a.

The wake is found to be unsteady in all the experiments. In particular, coherent oscillations centered at the neck region are visually evident from the schlieren data. The qualitative nature of the oscillations is found to be very similar to the observations of \citet[][henceforth referred to as SS2015]{schmidt2015oscillations}. The standard deviation map of temporal fluctuations in schlieren intensity from run 5 shown in figure~\ref{fig:cyl_inst_std}b highlights flow regions with high levels of unsteadiness. To extract a time scale for the oscillations, the modal analysis technique of spectral proper orthogonal decomposition (SPOD) \cite{towne2018spectral} was applied to schlieren data. SPOD is briefly described below prior to presentation of results.

SPOD is the frequency domain form of proper orthogonal decomposition (POD), and is suited for extracting spatio-temporal coherent structures in statistically stationary flows. The distinction between SPOD and standard POD is that SPOD modes are orthogonal under a space-time inner product, unlike POD modes that are only spatially orthogonal. Consider a time series consisting of $N$ snapshots sampled at frequency $F_s$ broken into $L$ blocks/realizations, each containing $M$ snapshots. Let $q_j^{\left(k\right)}$ be the $k^{\textrm{th}}$ realization of the vector of observations at discrete time $t_j$. The discrete Fourier transform for each realization is written as
\begin{equation}
    \widehat{q}^{\left(k\right)}_m = \sum_{j=0}^{M-1}q^{\left(k\right)}_{j+1}\mathrm{e}^{-\mathrm{i}2\pi jm/M} \;\;\;\;\mathrm{for}\;\;\;\; m=-\frac{M}{2}+1,...,\frac{M}{2}\, .
\end{equation}
The corresponding power spectral density (PSD) for each realization, denoted by $\widehat{P}^{\left(k\right)}_m$, then follows as 
\begin{equation}
    \widehat{P}^{\left(k\right)}_m = \frac{1}{M^2}\left|\widehat{q}^{\left(k\right)}_m\right|^2 \, .
\end{equation}
Subsequent analysis is performed at individual frequencies. A data matrix $\bm{\widehat{Q}^{\left(m\right)}}$ of size $P\times L$ corresponding to any one frequency is constructed such that each column corresponds to the PSD vector along $P$ spatial points. The cross-correlation matrix, $\bm{\widehat{S}^{\left(m\right)}}$, is written as
\begin{equation}
    \bm{\widehat{S}^{\left(m\right)}} = \bm{\widehat{Q}^{\left(m\right)}}\bm{\widehat{Q}^{\left(m\right)\,\mathrm{T}}}\, ,
\end{equation}
and the eigenvalue decomposition of the cross-correlation matrix is carried out as
\begin{equation}
    \bm{\widehat{S}^{\left(m\right)}\Phi^{\left(m\right)}} = \bm{\Phi^{\left(m\right)}\Lambda^{\left(m\right)}}\, .
\end{equation}
Here $\bm{\Phi^{\left(m\right)}}$ is a size $P\times P$ matrix whose columns $\phi^{\left(m\right)}_p$ $\left(p=1,2,...,P\right)$ are the SPOD modes at a particular frequency $f_m = mF_s/M$, and $\bm{\Lambda^{\left(m\right)}}$ is the corresponding diagonal matrix whose elements are the eigen values $\lambda^{\left(m\right)}_p$. The frequency is non-dimensionalized by defining a Strouhal number $St_D$ as
\begin{equation}
    St_D = \frac{fD}{U_{\infty}}.
\end{equation}
It is noted here that SS2015 extracted the oscillation Strouhal number by identifying a peak in the PSD of local fluctuations in shadowgraph intensity at a location in the neck region. In comparison, the present modal decomposition approach provides a global time scale for oscillations, and hence is better suited for flow characterization.

\begin{figure*}[t!]
    \centering
    \includegraphics[width=\textwidth]{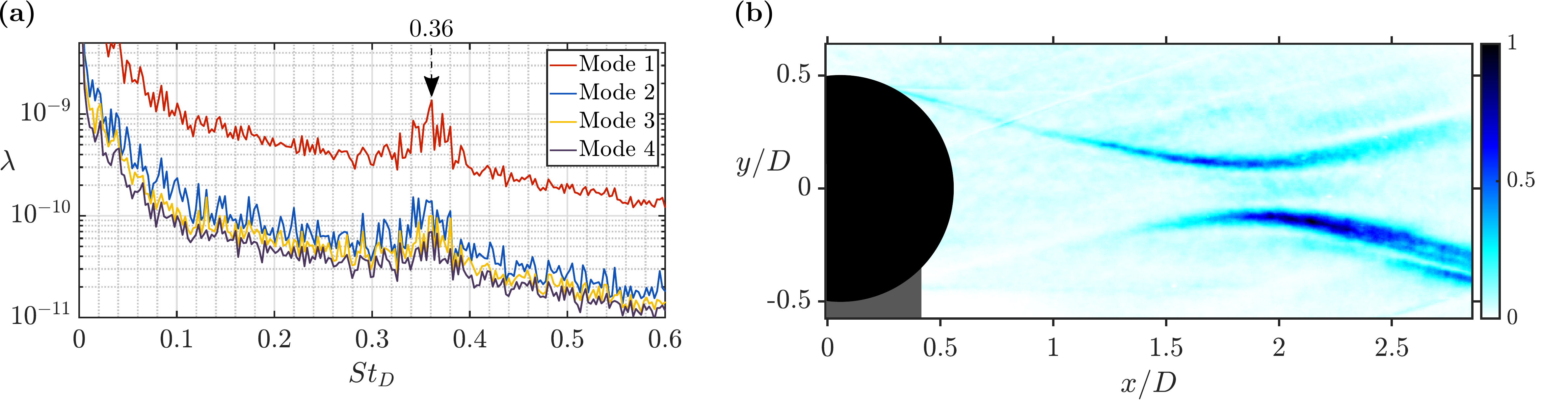}
    \caption{Results of spectral proper orthogonal decomposition for $Re_D = 4.3\times 10^5$. (a) Modal energy spectra of the first four modes. (b) Amplitude map of mode 1 at $St_D = 0.36$.}
    \label{fig:spod_results}
\end{figure*}

SPOD of schlieren data, with $L = 13$, $M = 2048$, $F_s = 10^5$, was performed for all the runs outlined in table~\ref{tab:exp_cond}. Figure \ref{fig:spod_results}a shows the modal energy spectra ($\lambda$) of the first four modes obtained from SPOD of data from run 5. The first mode is clearly seen to be more energetic in comparison with the other three modes, and hence any coherent spatial structures present in the flow are expected to be captured by this mode. Further, a peak in the spectrum is observed at $St_D = 0.36$ for mode 1, and spectra of other modes also reflect increased energy at the same Strouhal number. This clearly indicates $St_D = 0.36$ for coherent oscillations in the flow for run 5. Amplitude map of the corresponding mode 1, which represents the spatial coherent structure associated with oscillations, is shown in figure~\ref{fig:spod_results}b. The qualitative similarities between intensity maps shown figures~\ref{fig:cyl_inst_std}b and \ref{fig:spod_results}b indicates that oscillations with $St_D = 0.36$ dominate the overall flow unsteadiness. These SPOD results from run 5 are representative of results from the other five runs. 

\begin{figure*}
    \centering
    \includegraphics[width=\textwidth]{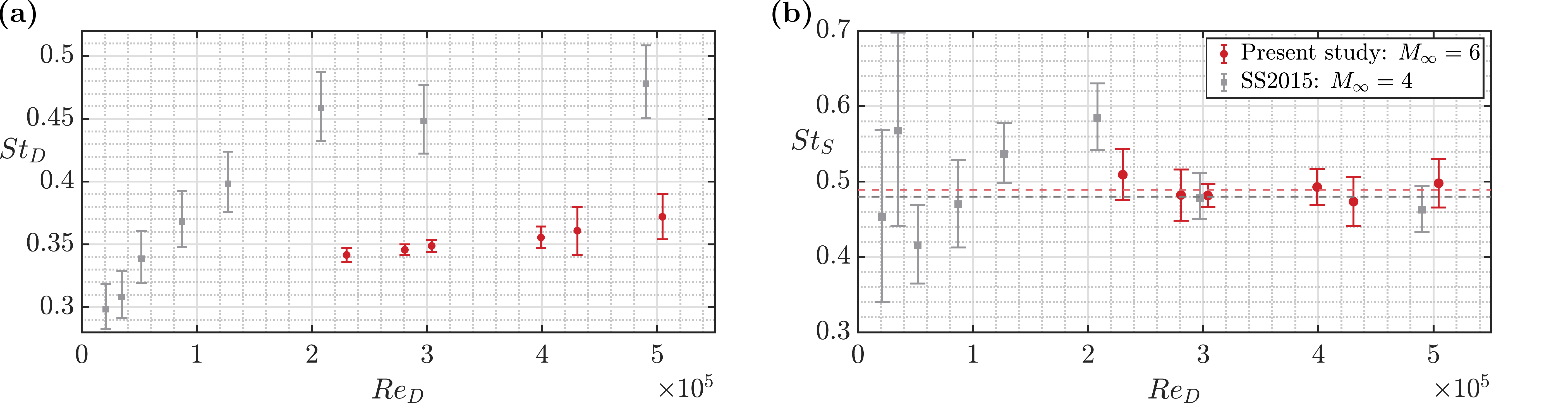}
    \caption{Strouhal numbers $St_D$ (a) and $St_S$ (b). The legend shown in (b) applies to both plots. The dashed lines in (b) indicate average values of $St_S$ for Mach 4 and Mach 6 data.}
    \label{fig:Strouhal_number}
\end{figure*}

The oscillation Strouhal number obtained from SPOD analysis for all six Reynolds numbers is shown in figure~\ref{fig:Strouhal_number}a along with Mach 4 data from SS2015. A systematic, though gradual, increase in $St_D$ with $Re_D$ is noted in the Mach 6 data. The influence of Mach number on $St_D$ is readily seen from the figure, with $St_D$ values from the two studies in the region of $Re_D$ overlap falling in two different ranges. Following SS2015, a Strouhal number $St_S$ based on the shear layer length $S$ (see figure~\ref{fig:cyl_schem}) is defined as
\begin{equation}
    St_S = \frac{fS}{U_{\infty}}.
\end{equation}
$St_S$ from the present experiments is shown in figure~\ref{fig:Strouhal_number}b along with Mach 4 data from SS2015. $St_S$ does not exhibit a systematic dependance on $Re_D$ and appears to be invariant in the considered range, with an average value of $St_S = 0.49$ across the six data points. There are two observations of interest and importance here. Firstly, this result corroborates Strouhal number universality with respect to Reynolds number reported by SS2015. Secondly, a close agreement between the $St_S$ value of 0.49 from the present study at Mach 6 and the $St_S$ value of 0.48 reported by SS2015 at Mach 4 shows that the universal behavior also extends to the Mach number, at least in the range $M_\infty$ = 4 to 6.

\begin{figure*}
    \centering
    \includegraphics[width=0.97\textwidth]{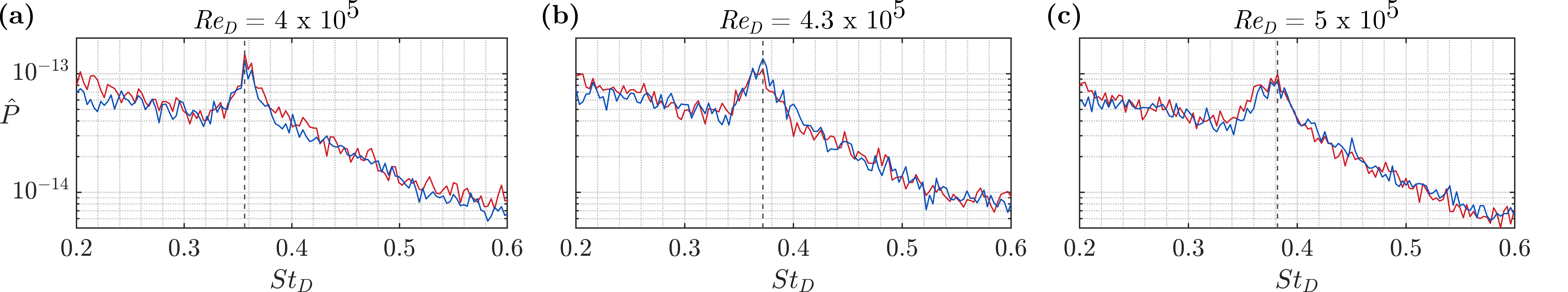}
    \caption{Power spectral density ($\hat{P}$) of non-dimensionalized pressure fluctuations $p'/P_{0}$ at the top  (\textcolor{firetruck}{\textbf{\textemdash}}) and bottom (\textcolor{sapphire}{\textbf{\textemdash}}) separation points. The dashed lines mark the peak $\hat{P}$ locations in $St_D$ for reference.}
    \label{fig:PSD_all}
\end{figure*}

\begin{figure}
    \centering
    \includegraphics[width=0.48\textwidth]{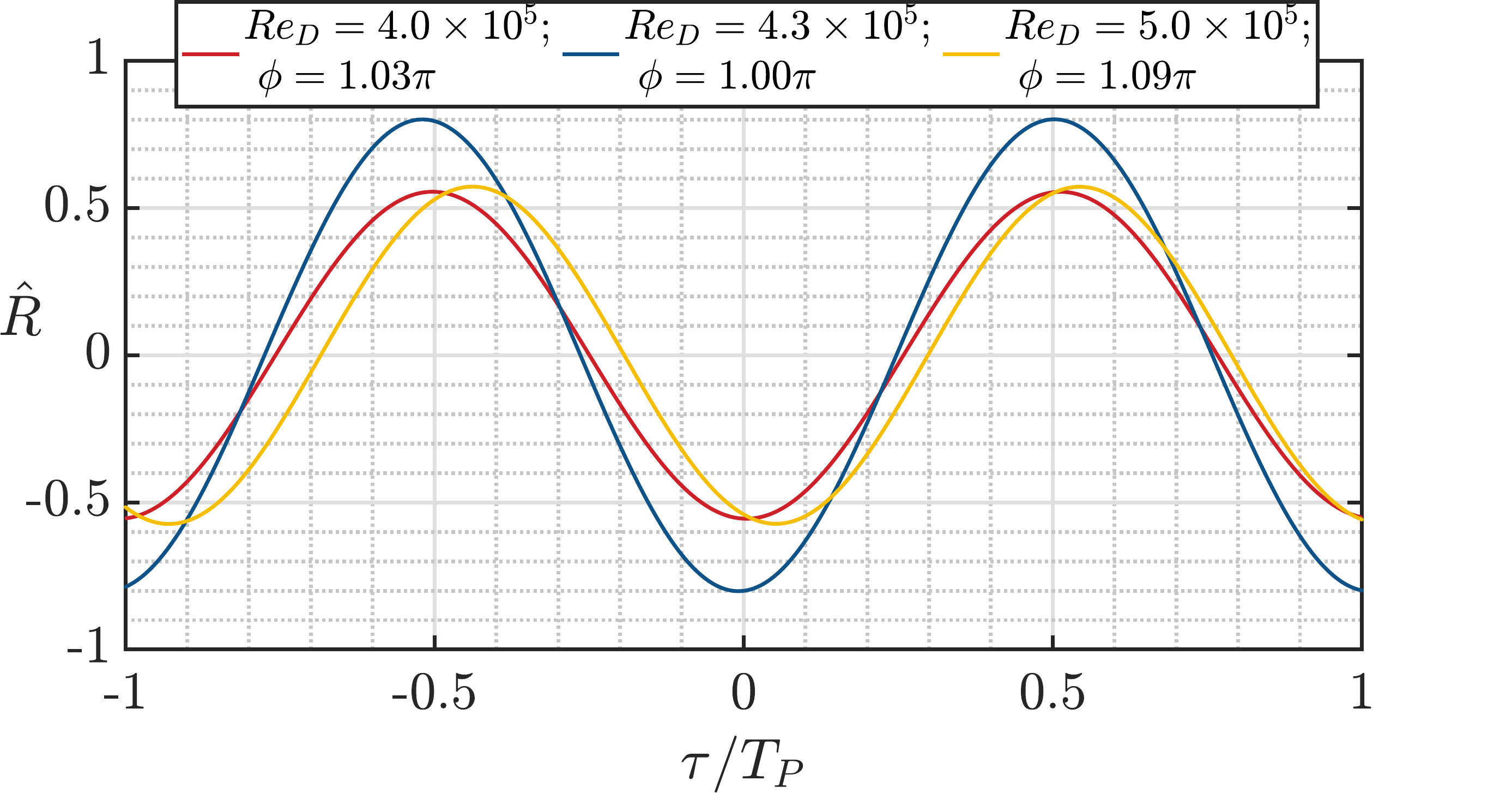}
    \caption{Cross-correlation coefficient between pressure fluctuations at the top and bottom separation points.}
    \label{fig:Corr_all}
\end{figure}

Based on the observation that the shear layer length yields a universal Strouhal number, SS2015 hypothesize that near-wake oscillations are sustained by communication between the neck and separation points through acoustic waves that propagate through the subsonic portions of the shear layers. From figure~\ref{fig:spod_results}b it is noted that a signature of the oscillation frequency is visible even in the flow region close to the separation point. PSD data of pressure fluctuations ($p'$) from the two transducers, which is shown in figure~\ref{fig:PSD_all}, confirms the same. At each of the three Reynolds numbers it is seen that peaks in the PSD curves from both sensors closely match the oscillation $St_D$ values obtained from SPOD analysis of corresponding schlieren data. Physical symmetry in the flow dictates that any oscillatory activity, with global coherence, at symmetric locations about the centerline ($y = 0$) should have a relative phase difference $\phi$ of either 0 or $\pi$ radians. The phase difference between the two pressure signals at each $Re_D$ is inferred by band-pass filtering them with a narrow band centered around the oscillation frequency, and then analysing the cross-correlation co-efficient $\hat{R}$ between the two band-passed signals. Band-pass filtering is done to mitigate the de-correlating effects of incoherent fluctuations from other parts of the power spectrum. Figure~\ref{fig:Corr_all} shows $\hat{R}$ as a function of time displacement (lead/lag) $\tau$ normalized by the oscillation time period $T_p = D U_{\infty}^{-1} St_{D}^{-1}$. At all three Reynolds numbers, peaks in $\hat{R}$ are found to be close to $\pm 0.5$, thereby indicating that oscillatory activity at one separation point is out of phase with the other. This implies that any coherent disturbances that propagate downstream along the length of the top and bottom shear layers are out of phase as they arrive at the neck, which perhaps is an important feature of the overall mechanism that governs the near-wake oscillations. Finally, it is noted that this aspect of the wake flow dynamics is the same as in the low-speed flow regime, where vortex shedding and accompanying coherent oscillations in other physical quantities in the flow have a $\pi$ radians phase offset between the two halves of the cylinder.

\section{BRIEF CONCLUSIONS \label{sec:conclusions}}

This experimental investigation of oscillations in the near-wake region of a 2D circular cylinder at Mach 6 confirms the universal behavior of Strouhal number (based on the shear layer length) with respect to Reynolds number reported earlier by SS2015. Further, the universal behavior is also seen to hold with respect to the flow Mach number. Pressure fluctuations at the two flow separation points on the cylinder aft surface indicate that coherent oscillatory activity in the two flow halves occurs with a phase offset of $\pi$ radians. The signature of oscillations seen in the surface pressure fluctuations lends some support to the hypothesis of SS2015 on the propagation of acoustics waves between the separation points and the neck. However, a detailed computation of the flow and stability analysis are desired toward obtaining a complete mechanistic understanding of the oscillations.


\bibliography{prfl2022}

\end{document}